\documentclass{aa}
\usepackage{txfonts}
\usepackage{natbib}
\bibpunct{(}{)}{;}{a}{}{,}
\usepackage[dvips]{graphicx}

\newcommand{\pun}[1]{\mbox{\rm\,#1}} % Command used to write physical units

\newcommand{\logg}{\ensuremath{\log g}}

\newcommand{\Teff}{\ensuremath{T_{\mathrm{eff}}}}

\newcommand{\beq}{\begin{equation}}
\newcommand{\eeq}{\end{equation}}

\newcommand{\eref}[1]{\mbox{(\ref{#1})}}
\newcommand{\ebref}[1]{\mbox{[\ref{#1}]}}
\newcommand{\plabel}[1]{\label{#1}}

\newcommand{\imu}{\ensuremath{m}}
\newcommand{\imup}{\ensuremath{m^\prime}}
\newcommand{\nmu}{\ensuremath{M}}
\newcommand{\ik}{\ensuremath{k}}
\newcommand{\ikp}{\ensuremath{k^\prime}}
\newcommand{\Nm}{\ensuremath{N_\imu}}

\newcommand{\Nphi}{\ensuremath{\tilde{N}_\imu}}

\newcommand{\Nphip}{\ensuremath{\tilde{N}_{\imu^\prime}}}
\newcommand{\Ntot}{\ensuremath{N}}
\newcommand{\Iobs}{\ensuremath{I}}
\newcommand{\Imu}{\ensuremath{I_{\imu}}}
\newcommand{\Ik}{\ensuremath{I_{\ik}}}
\newcommand{\Ikp}{\ensuremath{I_{\ikp}}}
\newcommand{\Imup}{\ensuremath{I_{\imup}}}
\newcommand{\Imuk}{\ensuremath{I_{\ik\imu}}}
\newcommand{\Imupkp}{\ensuremath{I_{\ikp\imup}}}
\newcommand{\HImu}{\ensuremath{\FT{I}_{\imu}}}

\newcommand{\HIk}{\ensuremath{\FT{I}_{\ik}}}

\newcommand{\HCIk}{\ensuremath{\conj{\FT{I}}_{\ik}}}
\newcommand{\HCIkp}{\ensuremath{\conj{\FT{I}}_{\ikp}}}

\newcommand{\HCImu}{\ensuremath{\conj{\FT{I}}_{\imu}}}

\newcommand{\HIobs}{\ensuremath{\FT{\Iobs}}}
\newcommand{\HCIobs}{\ensuremath{\conj{\FT{\Iobs}}}}

\newcommand{\mueff}{\ensuremath{\tilde{\mu}}}
\newcommand{\mueffmu}{\ensuremath{\mu_{\imu}}}
\newcommand{\mupeff}{\ensuremath{\mu_{\imup}}}
\newcommand{\muk}{\ensuremath{\mu_{\ik}}}
\newcommand{\mukp}{\ensuremath{\mu_{\ikp}}}

\newcommand{\LD}{\ensuremath{L}}
\newcommand{\LDcoef}{\ensuremath{a}}

\newcommand{\var}[1]{{\ensuremath{\sigma^2_{#1}}}}
\newcommand{\sig}[1]{{\ensuremath{\sigma_{#1}}}}
\newcommand{\cov}[2]{{\ensuremath{\mathrm{C}\left[#1,#2\right]}}}
\newcommand{\Fp}{\ensuremath{F}}
\newcommand{\Ap}{\ensuremath{A}}
\newcommand{\lp}{\ensuremath{l}}
\newcommand{\lgran}{\ensuremath{l_\mathrm{gran}}}
\newcommand{\fobs}{\ensuremath{f}}
\newcommand{\Rstar}{\ensuremath{R}}
\newcommand{\msum}[1]{\ensuremath{\sum_{#1=1}^{\nmu}}}

\newcommand{\nphisum}[1]{\ensuremath{\sum_{#1=1}^{\Nphi}}}
\newcommand{\nphipsum}[1]{\ensuremath{\sum_{#1=1}^{\Nphip}}}
\newcommand{\ufu}{\ensuremath{g}}
\newcommand{\wmu}{\ensuremath{w_\imu}}

\newcommand{\wmuk}{\ensuremath{w_{\ik\imu}}}
\newcommand{\wmupkp}{\ensuremath{w_{\ikp\imup}}}

\newcommand{\rpkkp}{\ensuremath{\phi_{\ik\ikp}}}

\newcommand{\blank}{\hspace{1ex}}
\newcommand{\tmean}[1]{\ensuremath{\left\langle #1\right\rangle}}
\newcommand{\emean}[1]{\ensuremath{\left\langle #1\right\rangle_\mathrm{ens}}}
\newcommand{\xmean}[1]{\ensuremath{\overline{#1}}}

\newcommand{\mom}[1]{\ensuremath{J_{#1}}}
\newcommand{\Hmom}[1]{\ensuremath{\FT{J}_{#1}}}
\newcommand{\HCmom}[1]{\ensuremath{\conj{\FT{J}}_{#1}}}
\newcommand{\Hfobs}{\ensuremath{\FT{\fobs}}}
\newcommand{\HCfobs}{\ensuremath{\conj{\FT{\fobs}}}}
\newcommand{\FT}[1]{\ensuremath{\hat{#1}}}
\newcommand{\conj}[1]{\ensuremath{#1^\ast}}

\newcommand{\xph}{\ensuremath{x_\mathrm{ph}}}
\newcommand{\Aint}{\ensuremath{\oint\!d\Ap}}

\newcommand{\Asum}[1]{\ensuremath{\sum_{#1=1}^{\Ntot}\Ap}}

\newcommand{\Ntotsum}[1]{\ensuremath{\sum_{#1=1}^{\Ntot}}}
\newcommand{\phiint}{\ensuremath{\int_{0}^{2\pi}\!\!\!d\varphi}}

\newcommand{\muint}{\ensuremath{\int_{0}^{1}\!\!d\mu}}

\newcommand{\Omint}{\ensuremath{\oint\!d\Omega}}

\newcommand{\abs}[1]{\ensuremath{\left|#1\right|}}
\newcommand{\xp}{\ensuremath{x^\prime}}
\newcommand{\yp}{\ensuremath{y^\prime}}
\newcommand{\delx}{\ensuremath{\Delta x}}
\newcommand{\dely}{\ensuremath{\Delta y}}
\newcommand{\NI}{\ensuremath{g}}
\newcommand{\NIxy}{\ensuremath{\NI(x,y)}}
\newcommand{\NIxpyp}{\ensuremath{\NI(\xp,\yp)}}
\newcommand{\NIxdxydy}{\ensuremath{\NI(x+\delx,y+\dely)}}
\newcommand{\Ixy}{\ensuremath{\Iobs(x,y)}}
\newcommand{\Ixpyp}{\ensuremath{\Iobs(\xp,\yp)}}
\newcommand{\xint}{\ensuremath{\int^{x_2}_{x_1}\!dx}}
\newcommand{\yint}{\ensuremath{\int^{y_2}_{y_1}\!dy}}

\newcommand{\draftflag}{false}

\begin{document}

\title{Hydrodynamical simulations of convection-related stellar micro-variability}
\subtitle{I. Statistical relations for photometric and photocentric variability}
% For referee layout:
\titlerunning{Hydrodynamical simulations \&\ stellar micro-variability}
%\titlerunning{}
%\authorrunning{H.-G. Ludwig}
\offprints{Hans-G\"unter Ludwig}

\author{Hans-G\"unter Ludwig\inst{1,2}}

\institute{Lund Observatory, Lund University, Box~43, 22100 Lund, Sweden
\and
GEPI, CIFIST, Observatoire de Paris-Meudon, 5 place Jules Janssen, 92195 Meudon
  Cedex, France\\ \email{Hans.Ludwig@obspm.fr}
}

\date{Received date; accepted date}

\abstract{%
Local-box hydrodynamical model atmospheres provide statistical information
  about the spatial dependence, as well as temporal evolution, of a star's
  emergent radiation field. Here, we consider late-type stellar atmospheres
  for which temporal changes of the radiative output are primarily related to
  convective (granular) surface flows.  We derived relations for evaluating
  the granulation-induced, disk-integrated thus observable fluctuations of the
  stellar brightness and location of the photocenter from radiation
  intensities available from a local model. Apart from their application in
  the context of hydrodynamical stellar atmospheres, these formulae provide
  some broader insight into the nature of the fluctuations under
  consideration.  Brightness fluctuations scale inversely proportional to the
  square root of the number of convective cells (the statistically
  independently radiating surface elements) present on the stellar surface and
  increase with more pronounced limb-darkening.  Fluctuations of the stellar
  photocentric position do \textit{not\/} depend on the number of cells and
  are largely insensitive to the degree of limb-darkening.  They amount to a
  small fraction of the typical cell size, and can become a limiting factor
  for high-precision astrometry in the case of extreme giants.  The temporal
  brightness and positional fluctuations are statistically uncorrelated but
  closely related in magnitude.
}

\maketitle

\section{Introduction}

The presence of granular flows on the surfaces of late-type stars implies
ultimate limits to their photometric stability, the stability of the position
of their photometric centroid (hereafter ``photocenter''), and the stability
of their spectroscopically determined radial velocities. The searches for
extrasolar planets and stellar p-mode oscillations have been pushing towards
higher observational sensitivity, now allowing the detection of the low-level
(hence ``micro-'') variability induced by the stochastically changing granular
flow pattern \citep[e.g.,][]{Kjeldsen+al99}. Stellar micro-variability becomes
particularly important for present and future space missions such as {\sc
  Most}, {\sc Corot}, {\sc Kepler}, {\sc Sim}, and {\sc Gaia}
\citep{Hatzes02,Green+al03,Aigrain+al04,Matthews+al04}. In this context,
granulation-related micro-variability is usually considered as a noise source.
It should be appreciated, however, that this ``noise'' also carries
information about the statistics of convective flows. Whether considered as
noise or signal: a better \emph{theoretical characterization of
  granulation-related micro-variability} would aid the planning and
interpretation of upcoming observations.

Based on detailed radiation-hydrodynamics simulations of convective surface
flows -- similar to the ones described in \citet{Wedemeyer+al04} -- we want to
predict the properties of granulation-related micro-variability in late-type
stars. To our knowledge the work of \citet{Trampedach+al98} is the only
example of such a theoretical effort to date. Trampedach et al. studied the
brightness and radial velocity variability in the Sun, $\alpha$~Cen~A, and
Procyon~A. Their hydrodynamical models, as well as the ones we refer to, are
so-called ``local-box'' models. They simulate convection in a small,
representative volume located within the stellar surface layers. In order to
derive observable, disk-integrated quantities, properties of the local model
have to be extrapolated to the full stellar disk.

This paper outlines the statistical procedure that we employ for that purpose.
It is intended as a precursor to later work presenting simulation results.
While the statistical developments are mostly straightforward, we nevertheless
feel that a somewhat detailed discussion is necessary to judge later results
and -- perhaps even more important -- their inherent uncertainties. In the
following we consider brightness fluctuations and fluctuations in the position
of the stellar photocenter. We do not consider the important case of
fluctuations in spectroscopic radial velocity, basically since related
spectral line formation issues complicate the problem. Despite its mostly
technical nature, this paper contains some results within a broader scope so
that the non-technically oriented reader might also find it of interest.

The paper is organized in the following way: in Sect.~\ref{s:assumptions} we
start by stating the basic statistical assumptions made when extrapolating
from the local simulations to the full stellar disk, and describe in
Sect.~\ref{s:modeloutput} what kind of simulation data did enter. In
Sect.~\ref{s:statpatch} we consider statistical properties of an individual
simulation patch, and in Sect.~\ref{s:statdisk} we continue by deriving the
related formulae for the expectation value, variance, and power spectra of
temporal brightness and positional variations of disk-integrated quantities.
We show that there is a general close connection between brightness and
positional variations, and that -- a somewhat counter-intuitive result --
photocentric variations do not depend on stellar radius but on the atmospheric
parameters alone.  Along the way, we introduce an approximation related to the
well-known Eddington approximation to mitigate some numerical shortcomings. In
Sect.~\ref{s:comments} we add some cautionary comments about short-cuts when
evaluating supposedly disk-integrated quantities from the flux data of a
simulation box. In Sect.~\ref{s:anamod} we discuss an illustrative analytical
model of brightness and positional variations, and make the connection between
simulation patches and convective cells. In Sect.~\ref{s:conclusions} we give
our conclusions.

\section{Basic statistical assumptions}
\plabel{s:assumptions}

Our goal was to make predictions of a star's brightness fluctuations and of
the displacements of its photocenter introduced by granulation, resorting to
results of radiation-hydrodynamics simulations. These simulations provide a
statistically representative, small, rectangular patch (or ``tile'') of the
emergent radiation field and its temporal evolution on the stellar surface. We
imagined the visible stellar hemisphere as tiled by a possibly large number of
simulation patches. Luckily, we did not have to construct this tiling, which
would have been difficult in practice due to the periodic lateral boundary
conditions employed in the simulations, among other problems.  To derive what
a distant observer would record if observing the unresolved stellar disk, we
made the following statistical and physical assumptions:
\begin{itemize}
\item Each patch radiates statistically independent of all other
  patches.
\item All patches on the stellar surface share the same statistical
      properties.
\item The space-wise statistics of the patches on the stellar surface
  can be replaced by the time-wise statistics of a simulated patch.
\end{itemize}

In a real star the first assumption is not exactly fulfilled. In particular,
acoustic modes have correlation lengths which introduce a coupling over scales
larger than are typically encompassed by a simulation box.  The coupling
between modes and convection is too weak to introduce a mode-related
long-range correlation into the convective part of the flow, so that the first
assumption should hold for granulation itself.  The modal part could be
treated separately as a perfectly correlated signal.  In supergiants one might
encounter limits to a statistical treatment of convection-related variability,
since only very few granules cover the stellar surface
\citep[e.g.,][]{Freytag03}.  In appendix~\ref{s:testindep} we comment on
testing the assumption of the statistical independence of a simulation patch.
The second assumption expresses the notion of homogeneity. Again, in a real
star, rotation or large-scale magnetic fields might lead to deviations from
homogeneity. The last assumption is a statement about ergodicity, which
appears justified considering the chaotic nature of stellar granular flows
\citep{Steffen+Freytag95}. All in all, we feel that the above assumptions are
not too critical in view of the precision we hope to achieve. Instead
limitations of the simulations themselves -- in particular the limited
statistics due to the finite length of the calculated time series -- control
the accuracy of predictions at the moment.

\section{The radiative output of a simulation patch}
\plabel{s:modeloutput}

In the hydrodynamical simulations, the radiative energy exchange is treated in
great detail: the geometry of the radiating flow is taken into account by a
ray-tracing technique, and the wavelength dependence by a number of wavelength
bands representing the effects of radiative heating or cooling in the
continuum and spectral lines. While we concentrate in the following on
wavelength integrated (``white light'') quantities, the formulas derived below
also hold for variables which are measured at a particular wavelength, or
which are integrals over a wavelength interval.

The basic radiative output that a simulation provides and that is relevant
here, are surface averages of the emergent intensity~\Imu\ for various
inclination cosines~$\mueffmu,\blank\imu\in\{1,\ldots,\nmu\}$; i.e. the
detailed spatial intensity information of a simulation is reduced to a level
familiar from 1D standard plane-parallel model atmospheres.  However, the
average intensities are computed for every time step, and are thus provided as
a function of time~$\Imu(t)$.

The discrete set of inclinations and azimuthal angles is rather small
reflecting limits imposed by the available computing resources.  Typically, we
worked with two or three angles in $\mu$- and four in $\varphi$-direction.
Table~\ref{t:muschemes} summarizes the employed inclinations and associated
weighting factors for evaluating the radiative flux. The integration scheme
with, in total, three inclinations~$M=3$ is a Gauss-Lobatto type integration
scheme \citep[e.g.,][]{Abramowitz+Stegun72}, while $M=2$ is not. The $M=2$
case is computationally quite economic for representing the center-to-limb
behavior in the flux calculation but has some disadvantages when it comes to
calculating the photocentric variability, as we discuss in detail later.

\begin{table}
\begin{center}
\caption[]{%
Inclination cosines~\mueffmu\ and weighting factors~\wmu\ for the flux
integration schemes with a total number of inclinations
of~$\nmu=2$ and $\nmu=3$. Factor $\wmu\mueffmu^2$ is given due to
its importance for evaluating some disk-integrated quantities.}
\label{t:muschemes}
\begin{tabular}{lccccc}
\hline\noalign{\smallskip}
\mbox{} & \multicolumn{2}{c}{$\nmu=2$} & \multicolumn{3}{c}{$\nmu=3$}\\
\noalign{\smallskip}
\hline\noalign{\smallskip}
\imu     & 1       & 2       & 1      & 2      & 3     \\
\mueffmu & 1.0000  & 0.3333  & 1.0000 & 0.7651 & 0.2852\\
\wmu     & 0.2500  & 0.7500  & 0.0667 & 0.3785 & 0.5549\\
$\wmu\mueffmu^2$ 
         & 0.2500  & 0.0833  & 0.0667 & 0.2216 & 0.0451\\
\noalign{\smallskip}
\hline
\end{tabular}%
\end{center}
\end{table}

\section{The statistics of the emergent flux of an individual
         simulation patch}
\plabel{s:statpatch}

In the following, we derive expressions both for the temporal average of the
emergent flux of a simulation patch and for its standard deviation.  Within
the discretization of the radiative transfer equation that is used in the
hydrodynamical simulations, the horizontal average of the emergent flux~\Fp\ 
(more precisely of its vertical component) of a simulation patch is given by
\beq 
\Fp \equiv \Omint\,\mu\Iobs = \muint\phiint\,\mu\Iobs
    \simeq 2\pi \msum{\imu}\nphisum{\ik}\wmuk\mueffmu\Imuk
\plabel{e:patchflux}
\eeq
%\begin{eqnarray} 
%\Fp & \equiv & \Omint\,\mu\Iobs = \muint\phiint\,\mu\Iobs\nonumber\\
%    & \simeq & 2\pi \msum{\imu}\nphisum{\ik}\wmuk\mueffmu\Imuk, 
%\plabel{e:patchflux}
%\end{eqnarray}
%
where \wmuk\ is the fraction of the solid angle of the hemisphere at
inclination~\mueffmu\ and at azimuthal angle $\varphi_{\ik}$ (see
Fig.~\ref{f:geometry}), and \Imuk\ the intensity in that direction.
Here and in the following, the symbol~``$\simeq$'' denotes the
approximate equality between continuous expressions and discrete
analogs, where the induced errors vanish in the limit of infinitely
fine discretization. We denote physical approximations
by~``$\approx$''.  Statistically, the convection pattern is
horizontally isotropic. Therefore, all weights belonging to the same
$\mu$-ring~\imu\ (for the definition see
Fig.~\ref{f:geometry}) are made equal since the discretization in
the azimuthal direction should reflect the isotropy. The weights differ
between different $\mu$-rings.  Introducing the total weight~\wmu\ of
the $\mu$-ring~\imu, the weights~\wmuk\ are normalized so that
\beq
\msum{\imu}\nphisum{\ik}\wmuk=\msum{\imu}\Nphi\wmuk=\msum{\imu}\wmu = 1 
\eeq
where $\Nphi=\wmu/\wmuk$ is the total number of solid angle elements belonging to
the $\mu$-ring~\imu. For later reference, we give the discrete analog
of the integral operator associated with the $\mu$-integration
\beq
\muint\,\ufu(\mu) \simeq \msum{\imu}\wmu \ufu_\imu
\plabel{e:mudispatch}
\eeq
where \ufu\ is an arbitrary function of~$\mu$ and
$\ufu_\imu=\ufu(\mueffmu)$. With~\eref{e:patchflux} we obtain for the
temporal average of the radiative flux~\Fp\ of a patch 
\beq
\tmean{\Fp} \simeq 2\pi\msum{\imu}\nphisum{\ik}\wmuk\mueffmu\tmean{\Imuk},
\eeq
since \wmuk\ and \mueffmu\ are time-independent.  Here and in the following,
\tmean{\ufu} denotes the temporal expectation value of a quantity~\ufu. Due to
the horizontal isotropy, all elements of the solid angle belonging to the same
$\mu$-ring share the same statistics
\beq
\tmean{\Imuk}=\tmean{\Imu} \blank\forall\:\ik,\imu\blank 
\eeq 
and
\beq
\sig{\Imuk}=\sig{\Imu} \blank\forall\:\ik,\imu\blank 
\eeq 
where $\sig{\ufu}$ denotes the standard deviation of a variable \ufu\
according to
\beq
\sig{\ufu} \equiv \sqrt{\tmean{\ufu^2}-\tmean{\ufu}^2}.
\eeq
Using the isotropy, we finally obtain for the temporal average of the
emergent flux
\beq
\tmean{\Fp} \simeq 2\pi\msum{\imu}\wmu\mueffmu\tmean{\Imu}.
\plabel{e:f}
\eeq

%\vspace{\baselineskip}
Now, we want to obtain an expression for the expectation value of~$\Fp^2$
introducing variances and correlation coefficients among the intensities going
into different directions of the solid angle.  The variances and correlation
coefficients could be evaluated from the data provided by the hydrodynamical
simulation. Of course, if one were interested in the flux variations of the
simulation alone, one could resort to the flux directly evaluated from
Eq.~\eref{e:patchflux} and would not -- rather indirectly -- turn to the
variances and correlation coefficients of the individual intensities.  We
derive the equations to compare them later with expressions for quantities
integrated over the stellar disk. Using the basic relation~\eref{e:patchflux}
we obtain
\beq
\tmean{\Fp^2}\simeq(2\pi)^2 \msum{\imu}\msum{\imup}\nphisum{\ik}
\nphipsum{\ikp}\wmuk\wmupkp\mueffmu\mupeff\tmean{\Imuk\Imupkp}.
\plabel{e:ff}
\eeq
%\begin{eqnarray}
%\lefteqn{\tmean{\Fp^2}=}\nonumber\\
%& &\!\!(2\pi)^2 \msum{\imu}\msum{\imup}\nphisum{\ik}
%\nphipsum{\ikp}\wmuk\wmupkp\mueffmu\mupeff\tmean{\Imuk\Imupkp}.
%\plabel{e:ff}
%\end{eqnarray}
%
We introduce the linear correlation coefficient between \Imuk\ and
\Imupkp\ \cov{\Imuk}{\Imupkp}, obeying the relation
\begin{eqnarray}
\tmean{\Imuk\Imupkp} 
& = & \tmean{\Imuk}\tmean{\Imupkp}+\sig{\Imuk}\sig{\Imupkp}\cov{\Imuk}{\Imupkp}
\nonumber\\
& = &\tmean{\Imu}\tmean{\Imup}+\sig{\Imu}\sig{\Imup}\cov{\Imuk}{\Imupkp}
\end{eqnarray}
with which we can rewrite expression~\eref{e:ff} as
\begin{eqnarray}
\lefteqn{\tmean{\Fp^2}\simeq(2\pi)^2\msum{\imu}\msum{\imup}\nphisum{\ik}
\nphipsum{\ikp}}\nonumber\\
&&\wmuk\wmupkp\mueffmu\mupeff
\left(\tmean{\Imu}\tmean{\Imup}+\sig{\Imu}\sig{\Imup}\cov{\Imuk}{\Imupkp}\right)
\nonumber\\
%& = & \left(2\pi\msum{\imu}\nphisum{\ik}
%  \tmean{\Imu}\wmuk\mueffmu\right)^2
%+\: (2\pi)^2\nonumber\\&&\msum{\imu}\msum{\imup}\nphisum{\ik}\nphipsum{\ikp}
%\sig{\Imu}\sig{\Imup}\cov{\Imuk}{\Imupkp}
%\wmuk\wmupkp\mueffmu\mupeff\nonumber\\
& = & \tmean{\Fp}^2 + (2\pi)^2
\msum{\imu}\msum{\imup}\nphisum{\ik}\nphipsum{\ikp}\nonumber\\
&&\wmuk\wmupkp\mueffmu\mupeff
\sig{\Imu}\sig{\Imup}\cov{\Imuk}{\Imupkp}.
\plabel{e:varf1}
\end{eqnarray}
To progress on the analytical side, we consider two
limiting and one intermediate case for the behavior among the
individual intensities emerging from a simulation patch:\\ 
i) The intensities of all elements of
the solid angle are uncorrelated, i.e.,
\beq
\cov{\Imuk}{\Imupkp}=\delta_{\imu\imup}\delta_{\ik\ikp}
\blank\blank\forall\imu,\imup,\ik,\ikp.
\eeq
ii) All intensities are perfectly positively correlated, i.e.,
\beq
\cov{\Imuk}{\Imupkp}=1
\blank\blank\forall\imu,\imup,\ik,\ikp.
\eeq
iii) The intensities are correlated for all azimuthal angles belonging
to the same inclination~\mueff, but are uncorrelated among
different inclinations, i.e.,
\beq
\cov{\Imuk}{\Imupkp}=\delta_{\imu\imup}
\blank\blank\forall\imu,\imup,\ik,\ikp.
\eeq
Case ii) often provides a good description of the actual behavior
found for the intensities of an individual patch. Case iii) is also a
plausible behavior, since different inclinations correspond to
different heights in the atmosphere; one would expect that first the
correlation between different height levels is lost before the
azimuthal correlation is reduced. In the three cases we obtain for
the variance of the flux of an individual patch
\begin{eqnarray}
\lefteqn{\var{\Fp}\simeq}\nonumber\\
& &(2\pi)^2\left\{
\begin{array}{lll}
\msum{\imu}\Nphi^{-1}\wmu^2\mueffmu^2\var{\Imu} & 
\mathrm{if}\blank
\cov{\Imuk}{\Imupkp}=\delta_{\imu\imup}\delta_{\ik\ikp}\\
\left(\msum{\imu}\wmu\mueffmu\sig{\Imu}\right)^2 &  
\mathrm{if}\blank \cov{\Imuk}{\Imupkp}=1\\
\msum{\imu}\wmu^2\mueffmu^2\var{\Imu} & 
\mathrm{if}\blank \cov{\Imuk}{\Imupkp}=\delta_{\imu\imup}.
\end{array}\right.
\plabel{e:varf2}
\end{eqnarray}
According to~\eref{e:varf1}, \sig{\Fp} is the sum of positive quantities so
that as relation among the cases we find
\beq
\sig{\Fp}^\mathrm{(ii)} \geq \sig{\Fp}^\mathrm{(iii)} \geq \sig{\Fp}^\mathrm{(i)},
\eeq 
reflecting the higher degree of cancellation for increasing degree of de-correlation.

\begin{figure}[t]
\begin{center}
\resizebox{0.7\hsize}{!}{\includegraphics[draft = \draftflag]%
{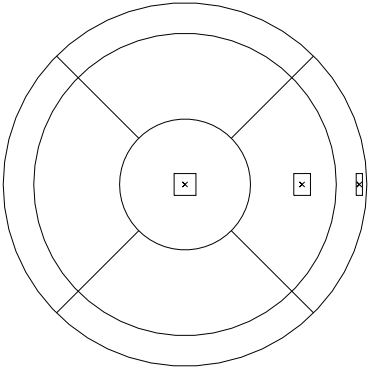}}
\vspace{-\baselineskip}
\caption[]{% 
Illustration of the weighting entering the calculation of the
radiative flux. The plot shows the projection of a hemisphere, the
concentric rings (in the text referred to as ``$\mu$-rings'') and
radial lines delineate the discrete elements of the solid angle
(respective surface area elements for the integration over the stellar
disk) entering the flux integration. Crosses mark the inclinations
used for representing the angular dependence of the radiation
field. The rectangles show how a square-shaped patch positioned on the
surface at the various inclination angles appears foreshortened
along the line of sight towards the limb. The diagram depicts the case $M=3$ of
Table~\ref{t:muschemes}.
\plabel{f:geometry} } % end of caption
\end{center}
\end{figure}

\section{Statistics of the disk-integrated, observable quantities}
\plabel{s:statdisk}

We now want to derive expressions for the brightness and positional
fluctuations that a distant observer would record when observing the
unresolved stellar disk.

\subsection{General statistical relations}

We are confronted with two problems: first, we have to introduce a
correlation length on the surface of the sphere describing the
two-point spatial correlation of the temporal granular intensity
signal. This we have done -- as outlined before -- by considering the
stellar surface as composed of identical patches which are
statistically independent from one another but share similar statistical
properties. We replace the gradual behavior of the
correlation function by a step-like behavior which simplifies the
treatment of the problem tremendously.  Second, we have to make best
contact to the data that the simulations provide. The central point
is how we describe the center-to-limb variation. The basic idea
is to transfer the discretization of the solid angle used in the
hydrodynamical simulation to a corresponding partitioning of the
stellar disk; i.e., we interpret the discretization of the solid angle
from Table~\ref{t:muschemes} as a partitioning of the visible stellar
surface, as depicted in Fig.~\ref{f:geometry}. This is of course
analogous to the way the flux of a standard plane-parallel
atmosphere model is translated to the disk integrated flux.  Since we
use a similar discretization of the stellar surface as in the case of
integration over solid angle, the statistical formulae describing
the flux of an individual simulation patch and disk-integrated
emission will look almost identical. This is intentional. However, as
will become apparent below, there are a number of subtleties which
deserve some attention and may affect the final outcome substantially.

Algebraically, we proceed by starting with the continuous description of a
quantity, and discretize into patches to introduce the finite correlation
length. We then return to the continuous description, since the analytical
manipulations are most easily done in this representation. To finally
establish contact with the simulation data, we discretize again applying the
scheme used in the hydrodynamical model.

We start by listing the discrete and continuous versions of the
involved integral operators.  Using the patches as discrete elements,
the surface integral of a function $\ufu(\mu=\cos\vartheta,\varphi)$ over
the stellar hemisphere can be written as
\beq
\Aint\,\ufu(\mu,\vartheta) = \muint\phiint\,\Rstar^2 \ufu(\mu,\vartheta) \simeq \Asum{\ik}\,\ufu_\ik
\plabel{e:Adis}
\eeq
where $\ufu_\ik=\ufu(\mu_\ik,\varphi_\ik)$, and $\mu_\ik$ and $\varphi_\ik$
are the coordinates of patch~$\ik$. The star's radius is \Rstar, and 
\Ntot\ is the total number of patches on the curved surface of the visible stellar
hemisphere; all patches have the same surface area
$\Ap=\Rstar^2\Delta\mu\,\Delta\varphi$, so that
\beq
\Ntot\Ap=2\pi\Rstar^2.
\plabel{e:ntota}
\eeq
For formal reasons we relate the surface tiling to a spherical
coordinate system, but in fact do not envision the stellar surface
tiled by segments suggested by such a system. The tiling should be
considered as locally close to isotropic --- as would be the case for
the tiling of a plane by squares. We repeat that an explicit
construction of a tiling turns out to be unnecessary.

We now formally define intensity-weighted integrals~\mom{\ufu} of an
arbitrary function~\ufu\ as
\beq
\mom{\ufu} \equiv \Aint\,\mu\Iobs\ufu.
\plabel{e:momdef}
\eeq
This definition is introduced to streamline our nomenclature. We shall
later set $\ufu=1$, and $\ufu=x$ where $x$ is the Cartesian $x$-position of
a patch on the stellar disk. \mom{1}\ is related to the
stellar flux, and \mom{x}\ to the position of the stellar photocenter.

Similar to the angular integration we now make assumptions about the
statistics of the patches. To preserve axisymmetric conditions, the
temporal expectation of the intensity should be a function of $\mu$
alone
\beq
\tmean{\Iobs}(\mu,\varphi)=\tmean{\Iobs}(\mu).
\plabel{e:intcon}
\eeq 
Analogously, the standard deviation of the intensity should obey
\beq
\sig{I}(\mu,\varphi)=\sig{I}(\mu).
\plabel{e:sigicon}
\eeq

\vspace{\baselineskip}

\subsubsection{Temporal expectation value of \mom{\ufu}}

In the following we consider a time-independent function
$\ufu(\mu,\varphi)$.  With the previous assumptions about the
$\mu$-dependence of the intensity and its variance, we obtain for the
temporal expectation value of \mom{\ufu}
\beq
\tmean{\mom{\ufu}} = \muint\phiint\,\Rstar^2\mu\tmean{\Iobs}\ufu
                = 2\pi\Rstar^2\muint\,\mu\tmean{\Iobs}\ufu,
\plabel{e:jvmean}
\eeq
where the last equality holds if \ufu\ is also $\varphi$-independent.

\subsubsection{Temporal variance of \mom{\ufu}}

The temporal expectation of $\mom{\ufu}^2$ depends on the properties of
the spatial correlation of the intensity, similar to the flux of a
  simulation patch as expressed by Eq.~\eref{e:ff}. 
We now introduce the discretization into patches~\eref{e:Adis},
which allows us to express these properties easily. 
The accuracy of this rests on the assumption that the patch size is small enough in
comparison to the global stellar scale that position-dependent quantities
(like the time-averaged intensity and \ufu) vary little across the patch,
but large enough that spatial correlations among statistical quantities
across patches are negligible. We obtain
\begin{eqnarray}
\tmean{\mom{\ufu}^2} &\simeq& 
\Ap^2\Ntotsum{\ik}\Ntotsum{\ikp}\muk\mukp\tmean{\Ik\Ikp}\ufu_\ik \ufu_{\ikp}\nonumber\\
 &=&\tmean{\mom{\ufu}}^2 + \Ap^2 \Ntotsum{\ik}\Ntotsum{\ikp}\muk\mukp\sig{\Ik}\sig{\Ikp}\ufu_\ik
\ufu_{\ikp}\cov{\Ik}{\Ikp},
%\lefteqn{\tmean{\mom{\ufu}^2}=% 
%\Ap^2\Ntotsum{\ik}\Ntotsum{\ikp}\muk\mukp\tmean{\Ik\Ikp}\ufu_\ik \ufu_{\ikp}}\nonumber\\
%&& =\tmean{\mom{\ufu}}^2 + \Ap^2 \Ntotsum{\ik}\Ntotsum{\ikp}\muk\mukp\sig{\Ik}\sig{\Ikp}\ufu_\ik
%\ufu_{\ikp}\cov{\Ik}{\Ikp},
\end{eqnarray}
where the intensity \Ik\ is the surface average over patch~\ik\ in the
direction towards the observer. In the present context summation
index~\ik\ does not refer to a particular coordinate direction but merely
numbers the patches located on the stellar surface. This reflects the fact
that we discuss correlations among the intensities emitted by different
patches and not within one patch as considered in Sect.~\ref{s:statpatch}.
The discretization into patches is not related to the partitioning
of the stellar disk as depicted in Fig.~\ref{f:geometry}.
This is introduced at a
later stage.  We are mostly interested in the case where the individual
patches are statistically independent, i.e.,
$\cov{\Ik}{\Ikp}=\delta_{\ik\ikp}$. However, to enable comparison with the
formulae derived for the flux of an individual patch, we also consider the case
of perfect positive correlation, i.e., $\cov{\Ik}{\Ikp}=1$. For the case of
statistical independence, we obtain
\beq
\var{\mom{\ufu}}\simeq\Ap^2\Ntotsum{\ik}\muk^2\var{\Ik}\ufu_\ik^2 
\simeq \Ap\Rstar^2\muint\phiint\,\mu^2\var{\Iobs}\ufu^2.
\plabel{e:varjvgeneral1}
\eeq
Note, that relation~\eref{e:varjvgeneral1} becomes asymptotically exact
if the spatial correlation length of \var{\Iobs}\ goes to zero.
For a $\varphi$-independent \ufu, we get
\begin{eqnarray}
\var{\mom{\ufu}}
&=&2\pi\Rstar^2\Ap\muint\,\mu^2\var{\Iobs}\ufu^2
 = (2\pi)^2\Rstar^4\Ntot^{-1}\muint\,\mu^2\var{\Iobs}\ufu^2\nonumber\\
&\simeq&(2\pi)^2\Rstar^4\Ntot^{-1}\msum{\imu}\wmu\mueffmu^2\var{\Imu}\ufu_\imu^2
\plabel{e:varjv1}
\end{eqnarray}
where in the last equality we introduced the
segmentation~\eref{e:mudispatch}.

For the second case of a coherent behavior of the patches, again
specializing for the case of a $\varphi$-independent \ufu, we obtain
\beq
\var{\mom{\ufu}} =\left(2\pi\Rstar^2\muint\,\mu\sig{\Iobs}\ufu\right)^2
\simeq (2\pi)^2\Rstar^4\left(\msum{\imu}\wmu\mueffmu\sig{\imu}\ufu_\imu\right)^2.
\plabel{e:varjv2}
\eeq
%\begin{eqnarray}
%\var{\mom{\ufu}} &=&\left(2\pi\Rstar^2\muint\,\mu\sig{\Iobs}v\right)^2\nonumber\\
%& = & (2\pi)^2\Rstar^4\left(\msum{\imu}\wmu\mueffmu\sig{\imu}\ufu_\imu\right)^2.
%\plabel{e:varjv2}
%\end{eqnarray}
%
In summary, for $\ufu(\mu,\varphi)=\ufu(\mu)$ in the two cases we obtained 
\beq
\var{\mom{\ufu}}=
(2\pi)^2\Rstar^4\left\{ 
\begin{array}{ll}
\Ntot^{-1} \muint\,\mu^2\var{\Iobs}\ufu^2
& \mathrm{if}\blank
\cov{\Ik}{\Ikp}=\delta_{\ik\ikp}\\
\left(\muint\,\mu\sig{\Iobs}\ufu\right)^2
&  \mathrm{if}\blank \cov{\Ik}{\Ikp}=1,
\end{array}\right.
\plabel{e:varjvana}
\eeq
%\begin{eqnarray}
%\lefteqn{\var{\mom{\ufu}}=(2\pi)^2\Rstar^4}\nonumber\\
%& &\cdot\left\{ 
%\begin{array}{ll}
%\Ntot^{-1} \muint\,\mu^2\var{\Iobs}\ufu^2
%& \mathrm{if}\blank
%\cov{\Ik}{\Ikp}=\delta_{\ik\ikp}\\
%\left(\muint\,\mu\sig{\Iobs}v\right)^2
%&  \mathrm{if}\blank \cov{\Ik}{\Ikp}=1,
%\end{array}\right.
%\plabel{e:varjvana}
%\end{eqnarray}
%
or the discrete analog using~\eref{e:mudispatch}
\beq
\var{\mom{\ufu}}\simeq(2\pi)^2 \Rstar^4
\left\{ 
\begin{array}{ll}
\Ntot^{-1}\msum{\imu}\wmu\mueffmu^2\var{\Imu}\ufu_\imu^2
& \!\mathrm{if}\!\blank
\cov{\Ik}{\Ikp}=\delta_{\ik\ikp}\\
\left(\msum{\imu}\wmu\mueffmu\sig{\Imu}\ufu_\imu\right)^2
&  \!\mathrm{if}\!\blank \cov{\Ik}{\Ikp}=1.
\end{array}\right.
\plabel{e:varjv}
\eeq
%\begin{eqnarray}
%\lefteqn{\var{\mom{\ufu}}=(2\pi)^2 \Rstar^4}\nonumber\\
%& &\cdot\left\{ 
%\begin{array}{ll}
%\Ntot^{-1}\msum{\imu}\wmu\mueffmu^2\var{\Imu}\ufu_\imu^2
%& \mathrm{if}\blank
%\cov{\Ik}{\Ikp}=\delta_{\ik\ikp}\\
%\left(\msum{\imu}\wmu\mueffmu\sig{\Imu}\ufu_\imu\right)^2
%&  \mathrm{if}\blank \cov{\Ik}{\Ikp}=1.
%\end{array}\right.
%\plabel{e:varjv}
%\end{eqnarray}

At this point we would also like to emphasize a general feature of the
formulae derived here and in the following: one might (erroneously)
conclude from formula~\eref{e:varjv} that \var{\mom{\ufu}}\ depends on
the particular, and to some extent arbitrary, geometry of the simulation
patch since \Ntot\ and \sig{\Imu} depend on the patch size. However,
only their combination $\sig{\Imu}/\sqrt{\Ntot}$ enters the relations.
This combination is invariant to changes in the patch geometry, as long
as our basic assumption holds that a patch represents a statistically
independent surface element. Hence, \var{\mom{\ufu}}\ is in fact
independent of the particular patch geometry.

\subsubsection{Temporal power spectrum of \mom{\ufu}}

To characterize the statistics of \mom{\ufu}\ further by
its temporal power spectrum, we only
consider the case of statistically independent patches. In the
following, $\FT{y}(\omega)$ denotes the temporal Fourier component of a
time-dependent variable $y(t)$, and \conj{\FT{y}}\ the conjugate
complex. Taking the Fourier transform of both sides of the basic
definition~\eref{e:momdef}, we get for an Fourier component of \mom{\ufu} 
\beq
\Hmom{\ufu} = \Aint\,\mu \FT{\Iobs}\ufu\simeq \Asum{\ik} \muk\HIk \ufu_\ik
\plabel{e:ftjv}
\eeq
where we assumed that \ufu\ is time-independent. Accordingly, 
for a frequency component of the power spectrum we obtain
\beq
\Hmom{\ufu}\HCmom{\ufu} \simeq
\Ntotsum{\ik}\Ntotsum{\ikp}\Ap^2\muk\mukp\HIk\HCIkp \ufu_\ik \ufu_{\ikp}.
\plabel{e:powerjv1}
\eeq
This expresses the power of a frequency component for a
particular realization of a convection pattern and its evolution. We
now have to average over all possible realizations to estimate
the expectation value of the power. Denoting the ensemble average
as \emean{\ufu}, we obtain
\beq
\emean{\Hmom{\ufu}\HCmom{\ufu}} \simeq
\Ntotsum{\ik}\Ntotsum{\ikp}\Ap^2\muk\mukp\emean{\HIk\HCIkp} \ufu_\ik \ufu_{\ikp}.
\plabel{e:powerjv2}
\eeq
All realizations share the property that the amplitude of a particular
Fourier component~\HIk\ is the same. Among the realizations, the
\HIk-components only differ by a uniformly distributed random phase
factor $\exp(i\rpkkp)$. The difference between two uniformly
distributed random phases is again a uniformly distributed random
phase.  It follows that
\beq
\emean{\HIk\HCIkp} =\abs{\HIk}\abs{\HCIkp}\,\emean{e^{i\Delta\rpkkp}} 
=\HIk\HCIk\,\delta_{\ik\ikp}.
\plabel{e:powerjv3}
\eeq
%
%The last equality comes about by the fact that the sum in the average
%\emean{e^{i\Delta\rpkkp}} can be interpreted as the result of a random
%walk in the complex plain. 
%The expectation value is zero as long as we
%consider two different points on the stellar surface,
%i.e. $\ik\neq\ikp$. Only in the case that we consider the same
%location, i.e. $\ik=\ikp$, the phase difference is not random but
%always zero, and the ensemble average of $\exp(i\Delta\rpkkp)$ is one. 
The last equality comes about because for $\ik\neq\ikp$ the ensemble
average \emean{e^{i\Delta\rpkkp}} corresponds to an average over the unit
circle in the complex plane which is zero. For $\ik=\ikp$ the phase
difference is not random but always zero, and the ensemble average is one.
From now on we drop the subscript ``ens'' indicating the ensemble average, but
keep in mind that with respect to average power, angular brackets denote ensemble
averages. One should further keep in mind that the hydrodynamical models
provide only an estimate of the average power of the intensity $\HIk\HCIk$,
because only a finite time interval is simulated.

With the formula for the ensemble average of the power of the
intensity we obtain for the power of \mom{\ufu}
\begin{eqnarray}
\tmean{\Hmom{\ufu}\HCmom{\ufu}}
&\simeq&\Ntotsum{\ik}\Ap^2\muk^2\tmean{\HIk\HCIk} \ufu^2_\ik\nonumber\\
&\simeq&\Ap\Rstar^2\muint\phiint\,\mu^2\tmean{\HIobs\HCIobs} \ufu^2.
\plabel{e:powerjv4}
\end{eqnarray}
Similar to Eq.~\eref{e:varjvgeneral1} the above relation becomes
asymptotically exact, if the spatial correlation length of the intensity goes
to zero.
%\begin{eqnarray}
%\tmean{\Hmom{\ufu}\HCmom{\ufu}}
%& = &\Ntotsum{\ik}\Ap^2\muk^2\tmean{\HIk\HCIk} \ufu^2_\ik\nonumber\\
%& = &\Ap\Rstar^2\muint\phiint\,\mu^2\tmean{\HIobs\HCIobs} \ufu^2.
%\plabel{e:powerjv4}
%\end{eqnarray}
%
Assuming a $\varphi$-independent \ufu, we finally obtain
\begin{eqnarray}
\tmean{\Hmom{\ufu}\HCmom{\ufu}}
&=& (2\pi)^2\Rstar^4\Ntot^{-1}\muint\,\mu^2\tmean{\HIobs\HCIobs} \ufu^2\nonumber\\
&\simeq& (2\pi)^2\Rstar^4\Ntot^{-1}\msum{\imu}\wmu\mueffmu^2\tmean{\HImu\HCImu} \ufu^2_\imu.
\plabel{e:powerjv}
\end{eqnarray}
Note, that the above relation for the expectation value of a frequency
component of the power spectrum of \mom{\ufu} is consistent with the
relation for its variance~\eref{e:varjv}: the frequency integral of the
power (excluding zero frequency) equals the variance of the signal
as expressed by Parseval's theorem.

\subsection{Statistics of the observable flux}
\plabel{s:flux}

We now apply the previously derived formulae to evaluate
statistical properties of the observable flux. The flux~\fobs\
observed at a distance~$D$ from the star is related to the
moment~\mom{1} according to $\mom{1}=D^2\fobs$.
Due to this close connection we shall sometimes loosely refer to
\mom{1} as observable flux.

\subsubsection{The expectation value of the observable flux}

From Eq.~\eref{e:jvmean} setting $\ufu=1$ and the basic definition of
  the flux of an individual patch~\eref{e:patchflux}, we obtain for a
  $\varphi$-independent mean intensity~\tmean{\Imu}
%
%\begin{samepage}%
%\begin{eqnarray}
%\tmean{\mom{1}} & = &2\pi\Rstar^2\muint\,\mu\tmean{\Iobs} 
%= 2\pi\Rstar^2\msum{\imu}\wmu\mueffmu\tmean{\Imu}\nonumber\\
%& = &\Rstar^2 \tmean{F}, 
%\plabel{e:j1=f}
%\end{eqnarray}%
%\end{samepage}%
%
\beq
\tmean{\mom{1}} = 2\pi\Rstar^2\muint\,\mu\tmean{\Iobs} 
%\simeq 2\pi\Rstar^2\msum{\imu}\wmu\mueffmu\tmean{\Imu}
%\simeq \Rstar^2 \tmean{F}, 
=\Rstar^2 \tmean{F}, 
\plabel{e:j1=f}
\eeq
or  
\beq
\tmean{\fobs} = \frac{\Rstar^2}{D^2}\tmean{F}.
\plabel{e:meanfobs}
\eeq
The observed temporally averaged flux -- perhaps no surprise -- is the average
flux of an individual patch reduced by the distance-related factor
${\Rstar^2}/{D^2}$. At this point one might be tempted to ignore the fact that
Eq.~\eref{e:meanfobs} expresses a relation between the expectation values of
\fobs\ \ and \Fp, not the fluxes themselves, and might calculate a power spectrum
of the disk-integrated flux as a power spectrum of the flux of the simulation
patch, perhaps scaled by the factor $\Ntot^{-1}$. We discuss the quantitative
error of this approximate procedure in Sect.~\ref{s:comments}.

\subsubsection{The variance of the observable flux}

To compute the variance of the observable brightness fluctuations, we
need to obtain the second moment of \mom{1}. As in the case
of an individual simulation patch, we consider the two limiting cases
of statistical independence and perfect positive correlation. Applying
Eq.~\eref{e:varjv} results in
\begin{samepage}%
\begin{eqnarray}%
\lefteqn{\var{\fobs}=\frac{\var{\mom{1}}}{D^4}}\nonumber\\
&&\simeq(2\pi)^2 \left(\frac{\Rstar}{D}\right)^4\left\{ 
\begin{array}{ll}
\Ntot^{-1}\msum{\imu}\wmu\mueffmu^2\var{\Imu}
& \mathrm{if}\blank
\cov{\Ik}{\Ikp}=\delta_{\ik\ikp}\\
\left(\msum{\imu}\wmu\mueffmu\sig{\Imu}\right)^2
&  \mathrm{if}\blank \cov{\Ik}{\Ikp}=1.
\end{array}\right.
\plabel{e:varfobs}
\end{eqnarray}%
\end{samepage}%
%\begin{samepage}%
%\begin{eqnarray}%
%\lefteqn{\var{\fobs}=\frac{\var{\mom{1}}}{D^4}
%=(2\pi)^2 \left(\frac{\Rstar}{D}\right)^4}\nonumber\\
%&&\cdot\left\{ 
%\begin{array}{ll}
%\Ntot^{-1}\msum{\imu}\wmu\mueffmu^2\var{\Imu}
%& \mathrm{if}\blank
%\cov{\Ik}{\Ikp}=\delta_{\ik\ikp}\\
%\left(\msum{\imu}\wmu\mueffmu\sig{\Imu}\right)^2
%&  \mathrm{if}\blank \cov{\Ik}{\Ikp}=1.
%\end{array}\right.
%\plabel{e:varfobs}
%\end{eqnarray}%
%\end{samepage}%
%
We find -- perhaps again no surprise -- that in the case of
statistical independence the variance of the observable flux scales
inversely proportional to the number of patches on the visible stellar
surface. Relations~\eref{e:varfobs} are close to
the results for an individual patch~\eref{e:varf2}. We can make it
even closer by introducing the number of patches~\Nm\ located within
$\mu$-ring~$\imu$ of the stellar disk
\beq
\Nm = \wmu\Ntot,
\plabel{e:nmwmu}
\eeq
since all patches on the stellar surface have the same area.

Since the weights in the solid angle integration are not proportional
to the number of solid angle elements, Eq.~\eref{e:varf2} cannot, on the other
hand, be cast in the form of Eq.~\eref{e:varfobs}.
%A corresponding relation for the number of solid angle elements
%belonging to a $\mu$-ring in the solid angle integration of a
%simulation patch does {\em not\/} hold since the individual solid
%angle elements generally differ in size between different $\mu$-rings so
%that $\Nphi\neq\wmu\Nsegs$ where \Nsegs\ is the total number of solid angle
%elements. 
Introducing~\eref{e:nmwmu}
makes Eq.~\eref{e:varfobs} read
\beq
\var{\fobs}
\simeq(2\pi)^2 \left(\frac{\Rstar}{D}\right)^4\left\{ 
\begin{array}{ll}
\msum{\imu}\Nm^{-1}\wmu^2\mueffmu^2\var{\Imu}
& \mathrm{if}\!\blank
\cov{\Ik}{\Ikp}=\delta_{\ik\ikp}\\
\left(\msum{\imu}\wmu\mueffmu\sig{\Imu}\right)^2
&  \mathrm{if}\!\blank \cov{\Ik}{\Ikp}=1.
\end{array}\right.
\plabel{e:varfobs2}
\eeq
%\begin{samepage}%
%\begin{eqnarray}
%\lefteqn{\var{\fobs}
%=(2\pi)^2 \left(\frac{\Rstar}{D}\right)^4}\nonumber\\
%&&\cdot\left\{ 
%\begin{array}{ll}
%\msum{\imu}\Nm^{-1}\wmu^2\mueffmu^2\var{\Imu}
%& \mathrm{if}\blank
%\cov{\Ik}{\Ikp}=\delta_{\ik\ikp}\\
%\left(\msum{\imu}\wmu\mueffmu\sig{\Imu}\right)^2
%&  \mathrm{if}\blank \cov{\Ik}{\Ikp}=1.
%\end{array}\right.
%\plabel{e:varfobs2}
%\end{eqnarray}
%\end{samepage}%
%
Only in the (for convective flows) unrealistic situation of a perfect
correlation among the intensities of a patch and also among the intensities
across the stellar disk, one can directly relate disk-integrated
brightness fluctuations to the fluctuations of the flux of the patch
as $\sig{\mom{1}}=\Rstar^2\sig{\Fp}$ which is analogous to the
relation for the flux $\tmean{\mom{1}}=\Rstar^2\tmean{\Fp}$.  In all
other cases one has to resort to the intensities themselves. Even
if one chose to make all elements of the angular integration the
same size, and obtained a direct correspondence of the formulae
this way, one still could not use the flux of the patch to derive
the brightness fluctuation of the star as a whole. The
reason is found in the different degrees of correlation among the intensities
of a patch and across the disk. Within a patch the intensities going
in various directions usually show a substantial degree of correlation,
which is lacking for the intensities emerging from different locations
on the stellar disk. We discuss this aspect
quantitatively in Sect.~\ref{s:comments}.

\subsubsection{Temporal power spectrum of the observable flux}

To further characterize the statistics of the observable flux, we
use Eq.~\eref{e:powerjv} to write down its temporal power
spectrum for the case of statistically independent patches
\beq
\tmean{\Hmom{1}\HCmom{1}} \simeq
(2\pi)^2\Rstar^4\Ntot^{-1}\msum{\imu}\wmu\mueffmu^2\tmean{\HImu\HCImu}.
\plabel{e:j1power1}
\eeq
For the power spectrum of the observable, relative flux variations we obtain
\beq
\frac{\tmean{\Hfobs\HCfobs}}{\tmean{\fobs}^2} = 
\frac{\tmean{\Hmom{1}\HCmom{1}}}{\tmean{\mom{1}}^2}
\simeq\Ntot^{-1}\frac{\msum{\imu}\wmu\mueffmu^2\tmean{\HImu\HCImu}}%
{\left(\msum{\imu}\wmu\mueffmu\tmean{\Imu}\right)^2}.
\plabel{e:j1power2}
\eeq
We remind the reader that the angular brackets around frequency
components of power spectra denote ensemble averages.

\subsection{Statistics of the photocentric position}
\plabel{s:xph}

We consider the position of the stellar photocenter in Cartesian coordinates
$(x,y)$.  The origin of the Cartesian coordinate system is placed in the
geometrical center of the stellar disk. We again assume statistically
isotropic conditions with respect to disk center. $x$- and $y$-direction are
equivalent, and we only have to regard -- without loss of generality -- one
Cartesian coordinate, say $x$. The position of the photocenter~$\xph(t)$ is
given as the intensity-weighted mean of the $x$-positions of all light
emitting patches tiling the visible stellar surface according to
\beq
\xph\equiv\frac{\mom{x}}{\mom{1}}
= \frac{\mom{x}}{\tmean{\mom{1}}}
\left[1+{\cal O}\left(\frac{\sig{\mom{1}}}{\tmean{\mom{1}}}\right)\right].
\plabel{e:defpc}
\eeq
Since the relative fluctuations of the observable flux are small, we
neglect the higher order terms in the fluctuations of \mom{1} and work
with the second equality in Eq.~\eref{e:defpc}. This leaves us with a much
more convenient {\em linear\/} relationship between \xph\ and the
intensities. For obtaining the expectation value and variance of \xph\
we now have to evaluate the respective quantities of \mom{x}. We do
this only for the case of uncorrelated patches, since it is clear that
one would not obtain any motion of photocenter for the case of a
perfectly correlated action of all patches.

To apply the previously derived formulae, we set $\ufu=x$, in
spherical coordinates
\beq
x = \Rstar\sqrt{1-\mu^2}\cos\varphi.
\plabel{e:xspherical} 
\eeq

\subsubsection{Expectation value of the photocentric position}

From Eq.~\eref{e:jvmean} one immediately gets
\beq
\tmean{\mom{x}} = \muint\phiint\,\Rstar^3\mu\tmean{\Iobs}\sqrt{1-\mu^2}\cos\varphi
                = 0,
\plabel{e:jxmean}
\eeq
and
\beq
\tmean{\xph}=0,
\plabel{e:xphmean}
\eeq
since the $\varphi$-integral vanishes. 
The expectation value of the photocenter's position coincides
with the geometric center of the star. This is of course a consequence
of the fact that we -- in a statistical sense -- preserved
axisymmetry in the formulation of the problem. 

\subsubsection{The variance of the photocentric position}
\plabel{s:varxph}

To obtain the variance of the photocentric position, we have to
evaluate \tmean{\mom{x}^2}. From relation~\eref{e:varjvgeneral1},
valid for $\varphi$-dependent quantities, we get 
\begin{eqnarray}
\var{\mom{x}}
&=&\Ap\Rstar^4\muint\phiint\,\mu^2\var{\Iobs}(1-\mu^2)\cos^2\varphi\nonumber\\
%&=& \frac{1}{2}\Rstar^2 2\pi\Rstar^2\Ap\muint\,\mu^2\var{\Iobs}(1-\mu^2)
%\nonumber\\
%&=& \frac{1}{2}\Rstar^2 (2\pi)^2 \Rstar^4 \Ntot^{-1}\muint\,\mu^2\var
%{\Iobs}(1-\mu^2)\nonumber\\
&=& \frac{1}{2}\Rstar^2\var{\mom{1}}\left[1 
- \frac{\muint\,\mu^4\var{\Iobs}}{\muint\,\mu^2\var{\Iobs}}\right],
\plabel{e:varjx}
\end{eqnarray}
where we used the relation (see Eq.~\ebref{e:varjv1})
\beq
\var{\mom{1}} = (2\pi)^2\Rstar^4\Ntot^{-1}\muint\,\mu^2\var{\Iobs}.
\plabel{e:varj1ana}
\eeq
Equation~\eref{e:varjx} describes a relation between \mom{x}\ and \mom{1}. We
shall see later that the connection is even tighter than it might at first appear from
Eq.~\eref{e:varjx}, since it turns out that the term in angular
brackets is only weakly variable with the degree of
  limb-darkening of \var{\Iobs}. Now we can write the variance of \xph:
\beq
\var{\xph}=\frac{\var{\mom{x}}}{\tmean{\mom{1}}^2}
=\frac{\Ap}{4\pi}\frac{\muint\,\mu^2\var{\Iobs}}{\left(\muint\,\mu\tmean{\Iobs}\right)^2}
\left[1-\frac{\muint\,\mu^4\var{\Iobs}}{\muint\,\mu^2\var{\Iobs}}\right].
\plabel{e:varxph1}
\eeq 

Remarkably, the variance of the photocentric position is independent of
stellar radius (or equivalently total number of patches), as long as we assume
that the patch size~\Ap\ and its radiation properties are fixed.
Astrophysically, this means that a star's photocentric
variability\footnote{Note that we are discussing the absolute spatial
  variability of the photocenter, hence the stellar distance does not enter at this
  point.}  depends only on its atmospheric parameters (\Teff, \logg, and
chemical composition). However, when changing a star's radius at fixed surface
conditions by altering its mass, the relative importance of sphericity is
changed. This is a higher order effect but might bring a radius
dependence back in for objects with extended atmospheres.  The radius independence is
intrinsic to the particular 2D axisymmetric geometry of the problem. It can be
shown that the same does not hold, e.g., for a linear chain of patches. From a
practical point of view it is convenient that we do not have have to specify a
precise stellar radius when evaluating the variance of the displacement of the
photocenter.

We have not written down the discrete analogs of
Eqs.~\eref{e:varjx} and~\eref{e:varxph1} yet, because a numerical problem  
exists when evaluating integrals involving
higher powers of $\mu$. The integrals become inaccurate; this is
particularly the case for our integration scheme with $M=2$, which is
not a Gauss-Lobatto type integration. We now seek a way to mitigate
the problem. To illustrate the situation, we assume that the dependence
of the expectation value of the intensity on $\mu$ follows a linear
limb-darkening law~\LD\ of the form
\beq
\frac{\tmean{\Iobs}(\mu)}{\tmean{\Iobs}(\mu=1)}
= \LD(\mu) = 1 - \LDcoef + \LDcoef\mu
\plabel{e:LDim}
\eeq
where \LDcoef\ is the limb-darkening coefficient.  The numerical
results of the hydrodynamical simulation show that the standard
deviation of the intensity is often roughly proportional to the
intensity. Hence, it is reasonable to assume a similar behavior for
the standard deviation
\beq
\frac{\sig{\Iobs}(\mu)}{\sig{\Iobs}(\mu=1)}
= \LD(\mu).
\plabel{e:LDisig}
\eeq
Using the linear limb-darkening laws~\eref{e:LDim} and~\eref{e:LDisig},
the integrals in Eq.~\eref{e:varxph1} can be evaluated
analytically (see appendix~\ref{s:LDmuintegrals}), as well as
numerically. The results can be compared to estimate the error
introduced by the various integration schemes.

Figure~\ref{f:varicomp} shows the comparison between the exact and
numerical results for the integral $\muint\,\LD^2\mu^2$, which enters
the calculation of \var{\mom{1}} (Eq.~\ebref{e:varj1ana}). For a
typical limb-darkening at optical wavelength we find a sizeable but
tolerable integration error in the case $M=2$. Case $M=3$ gives almost
exact results. Consequently, we do not expect that the limited accuracy
of our integration schemes will affect the evaluation of the
photometric variability substantially.

\begin{figure}
\begin{center}
\resizebox{\hsize}{!}{\includegraphics[draft = \draftflag]%
{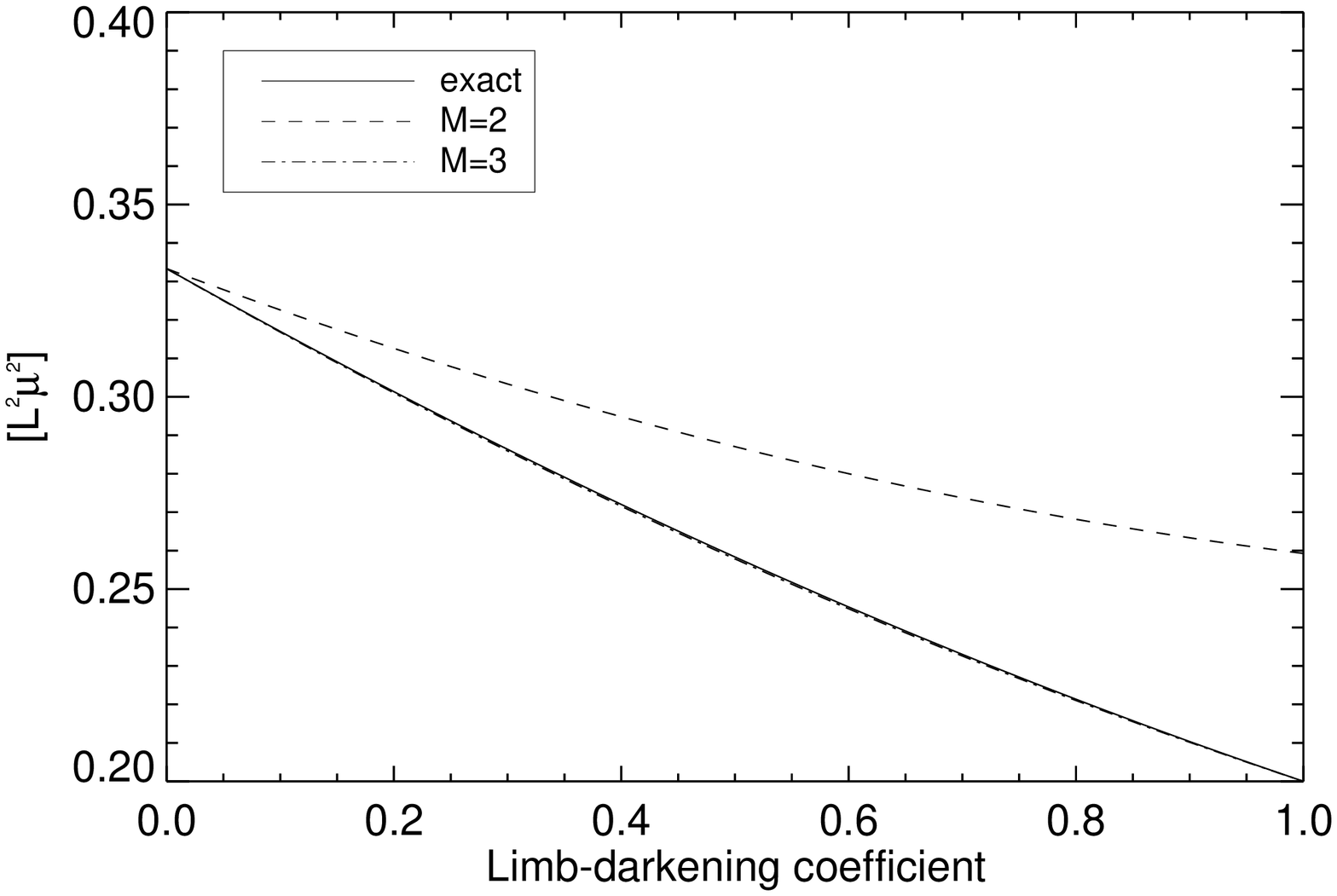}}
\caption[]{% 
Comparison of the integral $\muint\,\LD^2\mu^2$, which enters the
evaluation of \var{\mom{1}} (see Eq.~\ebref{e:varj1ana}). For a
typical limb-darkening around $\LDcoef=0.6$, we obtain a difference of
\mbox{13\,\%} to the exact result in the case $M=2$. In the case $M=3$,
differences are negligible.
\plabel{f:varicomp} 
} % end of caption
\end{center}
\end{figure}

As shown in Fig.~\ref{f:limbcomp}, the situation deteriorates when
considering the ratio $\muint\,\LD^2\mu^4/\muint\,\LD^2\mu^2$, which
enters the evaluation of the displacement of the photocenter
(Eq.~\ebref{e:varxph1}). However, the figure also shows that the
exactly evaluated overall variation of the ratio of the two integrals
is not large. In view of the large error associated with the numerical
integration in the case $M=2$, this leads to the idea of replacing the
ratio of the integrals by a constant, representative value of the
exact integral. We choose for the ratio
\beq
\frac{\muint\,\LD^2\mu^4}{\muint\,\LD^2\mu^2}\approx\frac{2}{3}, 
\plabel{e:myeddy}
\eeq
which results from a limb-darkening coefficient of about $\LDcoef=0.6$.
Apart from $\LDcoef=0.6$ being a
reasonable value for the limb-darkening coefficient in the optical, a
ratio of $\frac{2}{3}$ is chosen as a reminder that the
approximation~\eref{e:myeddy} is of course closely related to the
well-known Eddington approximation, where a value of $\frac{1}{3}$ is
assumed for the ratio between the second and zeroth angular moment of
the intensity. For this reason we shall refer to
approximation~\eref{e:myeddy} as the \emph{Eddington-like approximation}.

\begin{figure}
\begin{center}
\resizebox{\hsize}{!}{\includegraphics[draft = \draftflag]%
{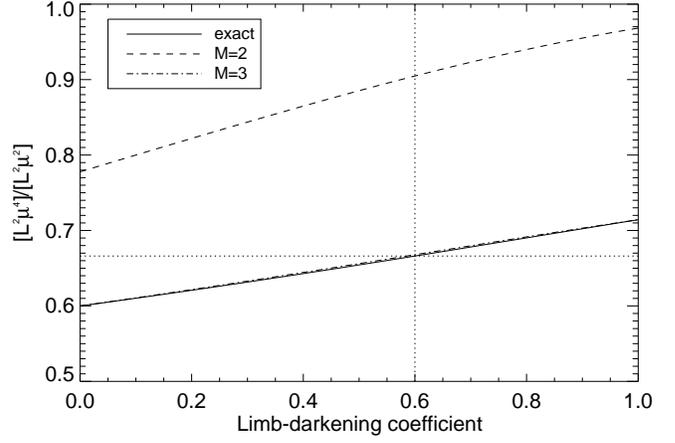}}
\caption[]{%
Comparison of $\muint\,\LD^2\mu^4/\muint\,\LD^2\mu^2$ which enters the
computation of the displacement of the photocenter (see
Eq.~\ebref{e:varxph1}). For a typical limb-darkening around
$\LDcoef=0.6$, we obtain a difference of \mbox{30\,\%} to the exact
result in the $M=2$ case. In the $M=3$ case, differences are
negligible. 
\plabel{f:limbcomp}
} % end of caption
\end{center}
\end{figure}

We apply the Eddington-like approximation to relations for the
displacement of the photocenter \eref{e:varjx}
and~\eref{e:varxph1}. In doing so we hope to mitigate the effects of
decreased numerical accuracy when evaluating the ratio
$\muint\,\LD^2\mu^4/\muint\,\LD^2\mu^2$. For the standard
deviation~\sig{\mom{x}}, we obtain
\beq
\sig{\mom{x}} \approx \frac{\Rstar}{\sqrt{6}}\sig{\mom{1}},
\plabel{e:varjxeddy}
\eeq
or for the standard deviation of the photocenter~\sig{\xph} itself,
\beq
\sig{\xph} \approx
\frac{\Rstar}{\sqrt{6}}\frac{\sig{\mom{1}}}{\tmean{\mom{1}}}
= \frac{\Rstar}{\sqrt{6}}\frac{\sig{\fobs}}{\tmean{\fobs}}.
\plabel{e:varxpheddy}
\eeq
Equation~\eref{e:varxpheddy} relates the observable flux variations to the
displacement of the photocenter. Remarkably, no information intrinsic to
granulation enters this relation. Hence, the relation could be more generally
valid than the particular derivation might suggest. And indeed, e.g.,
\citet{Lindegren77} has derived a similar connection in the context of the
astrometry of objects with an apparently time-variable surface pattern due to
rotation. Equation~\eref{e:varxpheddy} is a relation among quantities which
should be directly accessible to observation. This might be considered an
advantage. However, concerning convection as such, the diagnostic power of
Eq.~\eref{e:varxpheddy} is limited since the observed variations in brightness
and position cannot be immediately connected to convection. Other processes
can produce a similar signal.

Since the temporal power spectrum of the brightness fluctuations is closely
linked to its variance, Eq.~\eref{e:varxpheddy} also states that the power
spectrum of the photocentric displacement is -- up to the frequency
independent factor $\Rstar^2/6$ -- identical to the power spectrum of the
relative brightness fluctuations. This is a very convenient property. The
similarity between the power spectra of brightness and positional fluctuations
does not imply a temporal correlation among these variables. In fact, one can
immediately show that $\cov{\mom{1}}{\mom{x}}=0$ or, correspondingly, that the
observable flux and the displacement of the photocenter are uncorrelated:
\beq
\cov{\fobs}{\xph}=0.
\plabel{e:covfobsxph}
\eeq
\cite{Hatzes02} discusses simultaneous measurements of correlated observables
to improve signal-to-noise ratios in exoplanet searches. The observables
discussed here do not offer such a possibility.

Using the Eddington-like approximation in Eq.~\eref{e:varxph1},
we alternatively obtain 
\beq
\sig{\xph}
\approx\frac{\lp}{\sqrt{12\pi}}
\frac{\left(\muint\,\mu^2\var{\Iobs}\right)^\frac{1}{2}}%
{\muint\,\mu\tmean{\Iobs}}
\simeq\frac{\lp}{\sqrt{12\pi}}
\frac{\left(\msum{\imu}\wmu\mueffmu^2\var{\Imu}\right)^\frac{1}{2}}%
{\msum{\imu}\wmu\,\mueffmu\tmean{\Imu}},
\plabel{e:varxpheddy2}
\eeq
%\begin{eqnarray}
%\sig{\xph}
%&=&\frac{\lp}{\sqrt{12\pi}}
%\frac{\left(\muint\,\mu^2\var{\Iobs}\right)^\frac{1}{2}}%
%{\muint\,\mu\tmean{\Iobs}}\nonumber\\
%&=&\frac{\lp}{\sqrt{12\pi}}
%\frac{\left(\msum{\imu}\wmu\mueffmu^2\var{\Imu}\right)^\frac{1}{2}}%
%{\msum{\imu}\wmu\,\mueffmu\tmean{\Imu}}
%\plabel{e:varxpheddy2}
%\end{eqnarray}
%
where we have introduced the linear patch size
$\lp\equiv\sqrt{\Ap}$. Again, formula~\eref{e:varxpheddy2} is
manifestly radius-independent. It also gives the discrete expression
for evaluating the standard deviation of the photocentric
displacement.

\section{Estimating \sig{\fobs} from \sig{\Fp}?}
\plabel{s:comments}

In this section we briefly digress to discuss the relation between the flux
variations of a simulation patch as given by Eq.~\eref{e:varf2}, and the
disk-integrated flux variations given by Eq.~\eref{e:varfobs} more
quantitatively. The reason is that one might take the standard deviation of
the flux of a simulation patch scaled by $1/\sqrt{\Ntot}$ as a proxy for the
standard deviation of the observable flux. One might proceed in the same way
when calculating power spectra of the disk-integrated flux variations. For
deriving an estimate of the introduced errors, we once again assume the linear
limb-darkening laws~\eref{e:LDim} and~\eref{e:LDisig}. We evaluate the ratio
of the relative standard deviations scaled by the total number of patches
$\sqrt{\Ntot}(\sig{\fobs}/\fobs)/(\sig{\Fp}/{\Fp})$. We consider the cases ii)
and iii) (see Eq.~\ebref{e:varf2}) of the angular correlation of the
intensities of a patch, and compare those with the (realistic) case of a
statistically independent behavior of the patches across the stellar surface.
We restrict ourselves to cases ii) and iii), because they encompass the actual
range found for the angular intensity correlations within a patch.
Figure~\ref{f:pflux} depicts the result.  The exact result, i.e., analytically
integrated angular integrals, is left out in the case of partial correlation,
since no continuous analog of the discrete formula~\eref{e:varf2} (case iii)
can be easily written down. It is of minor importance since what is relevant
here are not the deviations from the exact result, but the absolute value of
the disk-to-patch ratios.

Figure~\ref{f:pflux} shows that one tends to {\em underestimate\/} the
amplitude of the observable brightness fluctuations, if one computes the
fluctuations directly from the flux of a simulation patch and simply scales it
by $1/\sqrt{\Ntot}$.  Depending on the angular correlation of the intensities
of the simulation patch, the error can become substantial. In the (not shown)
case of completely uncorrelated intensities of a simulation patch, the error
becomes even larger (a ratio of about~2 for $\nmu=2$, about~3.3 for $\nmu=3$,
both for four azimuthal angles) than in the cases depicted in
Fig.~\ref{f:pflux}. The two reasons for the discrepancies have already been
discussed in connection with Eqs.~\eref{e:varfobs} and~\eref{e:varfobs2}.

\begin{figure}
\begin{center}
\resizebox{\hsize}{!}{\includegraphics[draft = \draftflag]%
{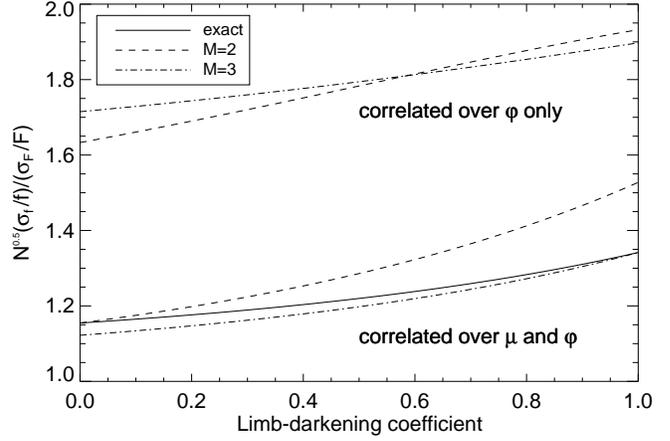}}
\caption[]{% 
The disk-to-patch ratio of relative flux variations as a function
of the limb-darkening coefficient for two different assumptions about the
angular correlation of the intensity within a patch. 
\plabel{f:pflux}
} % end of caption
\end{center}
\end{figure}

\section{An analytical model for the case of a linear limb-darkening law}
\plabel{s:anamod}

It is instructive to evaluate the formulae derived in
Sects.~\ref{s:flux} and~\ref{s:xph} for the linear limb-darkening
laws for \Iobs\ and \sig{\Iobs} (Eqs.~\eref{e:LDim}
and~\ebref{e:LDisig}). From Eqs.~\eref{e:jvmean}, \eref{e:varjv},
and auxiliary formulae from appendix~\ref{s:LDmuintegrals}, we directly
obtain for the relative fluctuations in the observable flux for the case of
uncorrelated patches
\beq
\frac{\sig{\fobs}}{\tmean{\fobs}} = \Ntot^{-\frac{1}{2}}
\frac{\sig{\Iobs}}{\tmean{\Iobs}} 
\frac{\left(\muint\,\LD^2\mu^2\right)^\frac{1}{2}}{\muint\,\LD\mu}
= \Ntot^{-\frac{1}{2}}\,\frac{\sig{\Iobs}}{\tmean{\Iobs}}\,
\frac{\sqrt{\frac{1}{3}-\frac{\LDcoef}{6}+\frac{\LDcoef^2}{30}}}{\frac{1}{2}-\frac{\LDcoef}{6}}.
\plabel{e:linearLD1}
\eeq
%\begin{eqnarray}
%\frac{\sig{\fobs}}{\tmean{\fobs}} &=& \Ntot^{-\frac{1}{2}}
%\frac{\sig{\Iobs}}{\tmean{\Iobs}} 
%\frac{\left(\muint\,\LD^2\mu^2\right)^\frac{1}{2}}{\muint\,\LD\mu}
%\nonumber\\
%&=& \Ntot^{-\frac{1}{2}}\,\frac{\sig{\Iobs}}{\tmean{\Iobs}}\,
%\frac{\sqrt{\frac{1}{3}-\frac{\LDcoef}{6}+\frac{\LDcoef^2}{30}}}{\frac{1}{2}-\frac{\LDcoef}{6}}.
%\plabel{e:linearLD1}
%\end{eqnarray}
%
Here \tmean{\Iobs} and \sig{\Iobs} should be understood as the values at
$\mu=1$. The \LDcoef-dependent numerical factor in
Eq.~\eref{e:linearLD1} has already been depicted in
Fig.~\ref{f:pflux}: the solid curve depicting the exact integral in
the case of complete correlation. Dependence on the
limb-darkening coefficient~\LDcoef\ is only mild, but it shows that
the brightness variations grow for a more pronounced limb-darkening.

Similarly, we obtain from Eq.~\eref{e:varxph1} for
the standard deviation of the photocentric displacement
\begin{eqnarray}
\sig{\xph} &=& \frac{\lp}{\sqrt{4\pi}}
\frac{\sig{\Iobs}}{\tmean{\Iobs}} 
\frac{\left(\muint\,\LD^2\mu^2(1-\mu^2)\right)^\frac{1}{2}}{\muint\,\LD\mu}\nonumber\\
&=& \frac{\lp}{\sqrt{4\pi}}\frac{\sig{\Iobs}}{\tmean{\Iobs}}\,
\frac{\sqrt{\frac{2}{15}-\frac{\LDcoef}{10}+\frac{\LDcoef^2}{42}}}%
{\frac{1}{2}-\frac{\LDcoef}{6}}
\plabel{e:linearLD2}
\end{eqnarray}
The \LDcoef-dependent factor in Eq.~\eref{e:linearLD2} is only
very weakly variable, as it varies (non-monotonically) between 0.201 and
0.206 with the limb-darkening coefficient in the interval
$\LDcoef\in[0,1]$ so that to good approximation
\beq
\sig{\xph}\approx 0.2\:\lp\,\frac{\sig{\Iobs}}{\tmean{\Iobs}}. 
\plabel{e:linearLD3}
\eeq
Formula~\eref{e:linearLD3} clearly emphasizes that the photocentric
displacement does not depend on the stellar radius but only on local,
granulation-related properties.

The chosen box-geometry of the hydrodynamical simulations sets the linear
 patch size, and the patch-averaged
 temporal intensity fluctuations scale with the surface area of the box. The
natural statistically independent elements on the stellar surface are
 not patches but the
granules, so that in physical terms one would like to connect $\lp$ to the
linear size of a granule and $\sig{\Iobs}/\tmean{\Iobs}$ to the time-wise
fluctuations of its emitted light. In the spirit of our basic assumption of
ergodicity, we might further replace the time-wise fluctuations by equivalent
space-wise fluctuations. We preliminarily find an approximate, empirical
relation
\beq
\lp\frac{\sig{\Iobs}}{\tmean{\Iobs}} \approx 0.4\:
\lgran\,\frac{\delta\Iobs_\mathrm{rms}}{\xmean{\Iobs}}, 
\plabel{e:stcontrast}
\eeq
which, combined with Eq.~\eref{e:linearLD3}, gives
\beq
\sig{\xph}\approx 0.08\:
\lgran\,\frac{\delta\Iobs_\mathrm{rms}}{\xmean{\Iobs}}.
\plabel{e:xphapprox}
\eeq
Here \lgran\ is the linear granular size -- defined via the maximum in the
spatial power spectrum of the intensity pattern -- \xmean{\Iobs} the
spatial average of the intensity, and $\delta\Iobs_\mathrm{rms}$ the
spatial root-mean-squares intensity contrast -- (conventionally)
defined at ideally infinite resolution, here the resolution of the
computational grid.  Equation~\eref{e:stcontrast} approximately
relates the patch-connected quantities in Eq.~\eref{e:linearLD3} to
physical properties of individual granules and the granulation
pattern. Perhaps surprising at first glance, the numerical factor in
Eq.~\eref{e:stcontrast} is not one but smaller than one, despite the
fact that the product $\lp{\sig{\Iobs}}/{\tmean{\Iobs}}$ should be
scale-invariant, as long as no spatial correlations are present on the
considered scales. The intensity contrast in
Eq.~\eref{e:stcontrast} refers to scales smaller than the granular
scale where this assumption is violated. On these scales the
intensities are spatially correlated leading to enhanced fluctuations
relative to larger scales, which needs to be compensated for by a
numerical factor smaller than one. Moreover, we also expect deviations
from one due to the differences in geometry between square-shaped
boxes and roundish granules.

As a rough estimate for the Sun from Eq.~\eref{e:xphapprox} with
$\lgran\approx 1000\pun{km}$ and
$\delta\Iobs_\mathrm{rms}/\xmean{\Iobs}\approx 0.18$, we obtain a granulation
induced photocentric variability of $\sig{\xph}\approx 14\pun{km}$, $\approx
10^{-7}\pun{AU}$, or $\approx
0.1\,\mu\mathrm{as\,/\,D}\left[\mathrm{pc}\right]$. This already indicates
that granulation will hardly affect the astrometry of dwarf stars at the
precision level that is expected to be reached by the {\sc Gaia} mission. The
size of granules grows roughly inversely proportional to the surface gravity
\citep{Freytag+al97} of a star, so that in giants and supergiants we expect
much greater effects.

\section{Conclusions and outlook}
\plabel{s:conclusions}

We derived formulae for evaluating the disk-integrated fluctuations of a
star's brightness and the position of its photocenter from local-box
hydrodynamical atmosphere models.  Besides the merely technical results, we
showed that there is a close connection between stellar photometric and
photocentric variability. While derived here for granular surface patterns,
the relation should be similar for any statistically homogeneous time-variable
surface pattern. Activity related brightness patterns are an interesting case.
Unfortunately, active regions are not expected to be homogeneously distributed
over the stellar surface but to follow belt-like zones. We nevertheless expect
that a semi-quantitative connection will still remain between activity-induced
photometric and photocentric variability so that order-of-magnitude estimates
of the photocentric variability are possible from measured brightness
fluctuations. This may help in estimating limitations to the achievable
astrometric accuracy for missions like {\sc Gaia}.  As we indicated in the
previous section, effects due to granulation themself are not likely to be
important with the exception of extreme giants. This will be quantified in
more detail in a subsequent paper. A preview of the simulation results can be
found in \cite{Svensson+Ludwig05}.
  
\acknowledgements 
We would like to thank Lennart Lindegren and Matthias Steffen for their help
concerning the algebra, statistical problems, and the presentation of the
results. This work benefitted from financial support by the Swedish Research
Council and the Royal Physiographic Society in Lund.

\appendix

\section{Angular integrals of $\LD^n\mu^m$}
\label{s:LDmuintegrals}

For a linear limb-darkening law of the form
\beq
\LD(\mu) = 1- \LDcoef + \LDcoef\mu,
\eeq
we find the following integrals over $\mu$:
\begin{eqnarray}
\muint\,\LD  \mu   \hfill   &=& \frac{1}{2}-\frac{\LDcoef}{6} 
                                \plabel{e:LDmu1}\\
\muint\,\LD^2\mu^2          &=& \frac{1}{3}-\frac{\LDcoef}{6}+\frac{\LDcoef^2}{30}
                                \plabel{e:LDmu2}\\
\muint\,\LD^2\mu^4          &=& \frac{1}{5}-\frac{\LDcoef}{15}+\frac{\LDcoef^2}{105}
                                \plabel{e:LDmu3}\\
\muint\,\LD^2\mu^2(1-\mu^2) &=& \frac{2}{15}-\frac{\LDcoef}{10}+\frac{\LDcoef^2}{42}.
                                \plabel{e:LDmu4}
\end{eqnarray}

\section{Statistical independence of a patch and the spatial
         autocorrelation of the granulation pattern}
\plabel{s:testindep}

One of our basic assumption is that the radiative output of a
simulated patch can be regarded as statistically independent of all
other patches.  We might check this assumption by testing whether the
assumption is already fulfilled within a simulation patch,
i.e., whether parts of the patch are already distant enough to radiate
independently. For this we have to evaluate the time-wise correlation
coefficient of the emitted intensity at two locations $(x,y)$ and
$(\xp,\yp)$ given by
\begin{eqnarray}
\lefteqn{\cov{\Ixy}{\Ixpyp}}\nonumber\\
&&=\frac{\tmean{\Ixy\Ixpyp}-\tmean{\Ixy}\tmean{\Ixpyp}}{\sig{\Ixy}\sig{\Ixpyp}}
%\nonumber\\
= \tmean{\NIxy\NIxpyp},
\end{eqnarray}
where we introduced the normalized intensity
\beq
\NI \equiv \frac{\Iobs-\tmean{\Iobs}}{\sig{\Iobs}}.
\eeq
Due to the horizontal statistical homogeneity of the granulation
pattern, the correlation is expected not to depend on the absolute
position but only on the distance between two points. To get an overall
measure of the statistical dependence~$C$ among the various locations,
one can spatially average over all pairs of points located a distance
$(\delx,\dely)$ apart:
\begin{eqnarray}
\lefteqn{C(\delx,\dely)}\nonumber\\
&&\!\!=\frac{1}{(x_2-x_1)(y_2-y_1)}\tmean{\xint\yint\,\NIxy\NIxdxydy}.
\end{eqnarray}
%\begin{eqnarray}
%\lefteqn{C(\delx,\dely)=\frac{1}{(x_2-x_1)(y_2-y_1)}}\nonumber\\
%%&&\cdot\xint\yint\,\tmean{\NIxy\NIxdxydy}\nonumber\\
%%&=&\frac{1}{(x_2-x_1)(y_2-y_1)}
%&&\cdot\tmean{\xint\yint\,\NIxy\NIxdxydy}.
%\end{eqnarray}
%
$C$ is given by the time-wise average of the autocorrelation function
of the normalized granular intensity divided by the considered surface
area. As stated before, $C$ depends on distance alone so that one
can further average the above expressions over circles
$\delx^2+\dely^2=\mathrm{const.}$, and obtain a one-dimensional
function describing the decrease of correlation with increasing
distance between two locations. In order to have our basic assumption
fulfilled, the spatial scale over which the correlation drops should be
significantly smaller than the linear patch size.

Here we have regarded the intensity, not the flux. However, since all
inclinations follow roughly the same behavior, the correlation length of the
intensities of all inclinations should be roughly the same. This implies a
corresponding correlation length for the radiative flux.

\bibliographystyle{aa}
\bibliography{2102manu}

\begin{thebibliography}{13}
\expandafter\ifx\csname natexlab\endcsname\relax\def\natexlab#1{#1}\fi

\bibitem[{Abramowitz \& Stegun(1972)}]{Abramowitz+Stegun72}
Abramowitz, M. \& Stegun, I. 1972, Handbook of mathematical functions (Dover
  Publications)

\bibitem[{Aigrain {et~al.}(2004)Aigrain, Favata, \& Gilmore}]{Aigrain+al04}
Aigrain, S., Favata, F., \& Gilmore, G. 2004, A\&A, 414, 1139

\bibitem[{Freytag(2003)}]{Freytag03}
Freytag, B. 2003, in Proceedings of the 12th Cambridge Workshop on Cool Stars,
  Stellar Systems, and the Sun, ed. A.~Brown, G.~Harper, \& T.~Ayres,
  1024--1029

\bibitem[{Freytag {et~al.}(1997)Freytag, Holweger, Steffen, \&
  Ludwig}]{Freytag+al97}
Freytag, B., Holweger, H., Steffen, M., \& Ludwig, H.-G. 1997, in Science with
  the VLT Interferometer, ed. F.~Paresce (Springer), 316--317

\bibitem[{Green {et~al.}(2003)Green, Matthews, Seager, \&
  Kuschnik}]{Green+al03}
Green, D., Matthews, J., Seager, S., \& Kuschnik, R. 2003, Astrophys. J., 597,
  590

\bibitem[{Hatzes(2002)}]{Hatzes02}
Hatzes, A. 2002, Astron. Nachr., 323, 392

\bibitem[{Kjeldsen {et~al.}(1999)Kjeldsen, Bedding, Frandsen, \&
  Dall}]{Kjeldsen+al99}
Kjeldsen, H., Bedding, T., Frandsen, S., \& Dall, T. 1999, MNRAS, 303, 579

\bibitem[{Lindegren(1977)}]{Lindegren77}
Lindegren, L. 1977, A\&A, 57, 55

\bibitem[{Matthews {et~al.}(2004)Matthews, Kuschnik, Guenther, Walker, Moffat,
  Rucinski, Sasselov, \& Weiss}]{Matthews+al04}
Matthews, J., Kuschnik, R., Guenther, D., {et~al.} 2004, Nature, 430, 51

\bibitem[{Steffen \& Freytag(1995)}]{Steffen+Freytag95}
Steffen, M. \& Freytag, B. 1995, Chaos, Solitons \&\ Fractals, 5, no.~10, 1965

\bibitem[{Svensson \& Ludwig(2005)}]{Svensson+Ludwig05}
Svensson, F. \& Ludwig, H.-G. 2005, in Proceedings of the 13th Cambridge
  Workshop on Cool Stars, Stellar Systems, and the Sun (ESA SP-560), ed.
  F.~Favata, G.~Hussain, \& B.~Battrick (ESA Publications Devision), 979--984

\bibitem[{Trampedach {et~al.}(1998)Trampedach, Christensen-Dalsgaard, Nordlund,
  \& Stein}]{Trampedach+al98}
Trampedach, R., Christensen-Dalsgaard, J., Nordlund, {\AA}., \& Stein, R. 1998,
  in The First MONS Workshop: Science with a Small Space Telescope, ed.
  H.~Kjeldsen \& T.~Bedding (Aarhus Universitet), 59

\bibitem[{Wedemeyer {et~al.}(2004)Wedemeyer, Freytag, Steffen, Ludwig, \&
  Holweger}]{Wedemeyer+al04}
Wedemeyer, S., Freytag, B., Steffen, M., Ludwig, H.-G., \& Holweger, H. 2004,
  A\&A, 414, 1121

\end{thebibliography}

\end{document}